%% file: main.tex
\documentclass[conference]{IEEEtran}
\usepackage{amsmath,amsfonts}
\usepackage{array}
\usepackage{textcomp}
\usepackage{stfloats}
\usepackage{url}
\usepackage{verbatim}
\usepackage{graphicx}
\usepackage{cite}
\usepackage{xcolor}
\usepackage[english]{babel}
\usepackage{booktabs}

\usepackage{algorithm}
\usepackage{algpseudocode}
\usepackage{amsmath}

\usepackage{subcaption}

\usepackage[acronym]{glossaries}
\makeglossaries
\loadglsentries{acronyms}

\usepackage{amsmath}
\usepackage{graphicx}

\usepackage{tikz}
\usetikzlibrary{fit,calc}

\colorlet{mypink}{red!40}
\colorlet{myblue}{cyan!60}

\makeatletter 
\newcommand{\linebreakand}{%
  \end{@IEEEauthorhalign}
  \par
  \mbox{}\hfill\begin{@IEEEauthorhalign}
}
\makeatother 

\title{Decentralized Spatial Reuse Optimization in Wi-Fi: An Internal Regret Minimization Approach}

\author{\IEEEauthorblockN{Francesc Wilhelmi}
\IEEEauthorblockA{Wireless Networking\\
Universitat Pompeu Fabra, 
Spain\\
Email: francisco.wilhelmi@upf.edu}
\and
\IEEEauthorblockN{Boris Bellalta}
\IEEEauthorblockA{Wireless Networking\\
Universitat Pompeu Fabra, 
Spain\\
Email: boris.bellalta@upf.edu}
\and
\IEEEauthorblockN{Miguel Casasnovas}
\IEEEauthorblockA{Wireless Networking\\
Universitat Pompeu Fabra, 
Spain\\
Email: miguel.casasnovas@upf.edu}
\linebreakand
\and
\IEEEauthorblockN{Aleksandra Kijanka}
\IEEEauthorblockA{Wireless Networking\\
Universitat Pompeu Fabra, 
Spain\\
Email: aleksandra.kijanka@upf.edu}
\and
\IEEEauthorblockN{Miguel Calvo-Fullana}
\IEEEauthorblockA{Wireless \& Secure Communications\\
Universitat Pompeu Fabra, 
Spain\\
Email: miguel.calvo@upf.edu}}

\begin{document}
\maketitle

\begin{abstract}
Spatial Reuse (SR) is a cost-effective technique for improving spectral efficiency in dense IEEE 802.11 deployments by enabling simultaneous transmissions. However, the decentralized optimization of SR parameters---transmission power and Carrier Sensing Threshold (CST)---across different Basic Service Sets (BSSs) is challenging due to the lack of global state information. In addition, the concurrent operation of multiple agents creates a highly non-stationary environment, often resulting in suboptimal global configurations (e.g., using the maximum possible transmission power by default). To overcome these limitations, this paper introduces a decentralized learning algorithm based on regret-matching, grounded in internal regret minimization. Unlike standard decentralized ``selfish'' approaches that often converge to inefficient Nash Equilibria (NE), internal regret minimization guides competing agents toward Correlated Equilibria (CE), effectively mimicking coordination without explicit communication. Through simulation results, we showcase the superiority of our proposed approach and its ability to reach near-optimal global performance. These results confirm the not-yet-unleashed potential of scalable decentralized solutions and question the need for the heavy signaling overheads and architectural complexity associated with emerging centralized solutions like Multi-Access Point Coordination (MAPC).
\end{abstract}


\IEEEpeerreviewmaketitle

\section{Introduction}

The proliferation of wireless devices and the exponential growth in traffic demands have pushed Wi-Fi to undergo an ambitious transformation and adopt significant architectural changes~\cite{geraci2025wi}. IEEE 802.11 relies on \gls{csmaca}, which has been proven effective in sparse networks, but performs poorly in dense deployments where neighboring \glspl{bss} share time resources inefficiently (e.g., by imposing conservative carrier sensing policies that lead to excessive contention). To address this, among many other features, recent amendments such as 802.11ax (Wi-Fi 6) and 802.11be (Wi-Fi 7) have introduced \gls{sr}, which allows devices to ignore inter-\gls{bss} interference below a certain \gls{cst} (technically referred to as \gls{obsspd} threshold), thereby enabling concurrent transmissions on the same frequency channels. By combining \gls{cst} adjustment and \gls{tpc}, \gls{sr} can potentially unlock more simultaneous transmissions and thus increase network performance. 

However, Wi-Fi's decentralized nature (i.e., \glspl{bss} operate autonomously, lacking information regarding the channel state, traffic load, or configuration of neighboring networks), hinders the optimization of \gls{sr}. When multiple uncoordinated \glspl{bss} attempt to concurrently optimize \gls{cst} and transmit power---whose relationship is highly non-linear and coupled---, they create a non-stationary environment, where the interactions between devices change constantly. Under these conditions, online learning approaches emerge as a compelling solution to achieve global \gls{sr} configurations~\cite{jamil2016novel}. Still, state-of-the-art decentralized learning algorithms, which typically seek to maximize individual utility (external regret minimization), may fail to find globally optimal and stable solutions, thus leading to inefficient \gls{ne}~\cite{wilhelmi2019collaborative}. 

Currently, the \gls{tgbn} is developing the next generation of the 802.11 standard, 802.11bn (Wi-Fi 8)~\cite{geraci2025wi, 11bn_draft}, where \gls{mapc} is proposed to address some of the decentralization challenges. While coordination can enable optimal configurations, it introduces significant scalability barriers, including substantial signaling overhead and synchronization requirements. This is the reason why, so far, \gls{mapc} limits coordination to a maximum of two \glspl{ap}. In this paper, we propose a decentralized, online learning alternative that achieves the performance benefits of coordination without the associated overhead. We introduce an algorithm based on regret-matching~\cite{hart2000simple} that aims to minimize the internal regret of agents. Unlike standard algorithms that compare a strategy’s performance against a fixed best alternative, internal regret minimization algorithms evaluate how much better an agent would have performed if it had swapped one specific action for another. By doing this, our algorithm guides independent \glspl{bss} toward a \gls{ce}---a state of implicit coordination that maximizes global welfare. Our contribution demonstrates that scalability and optimality are not mutually exclusive in next-generation Wi-Fi networks. We show that by engineering the learning objective to minimize internal regret, Wi-Fi devices can learn efficient \gls{sr} patterns, questioning the need for complex, centralized coordination architectures.

The remainder of this paper is structured as follows: Section~\ref{sec:related_work} reviews Wi-Fi's \gls{sr} and the related literature. Section~\ref{sec:formulation} formulates the \gls{sr} problem as a multi-player game and describes the proposed internal regret minimization solution, which is subsequently evaluated in Section~\ref{sec:performance_evaluation}. Finally, Section~\ref{sec:conclusions} provides concluding remarks.

\section{Learning Spatial Reuse in Wi-Fi}
\label{sec:related_work}

\subsection{Spatial Reuse in Wireless Communications}

\begin{figure}[t!]
    \centering
    \includegraphics[width=.85\columnwidth]{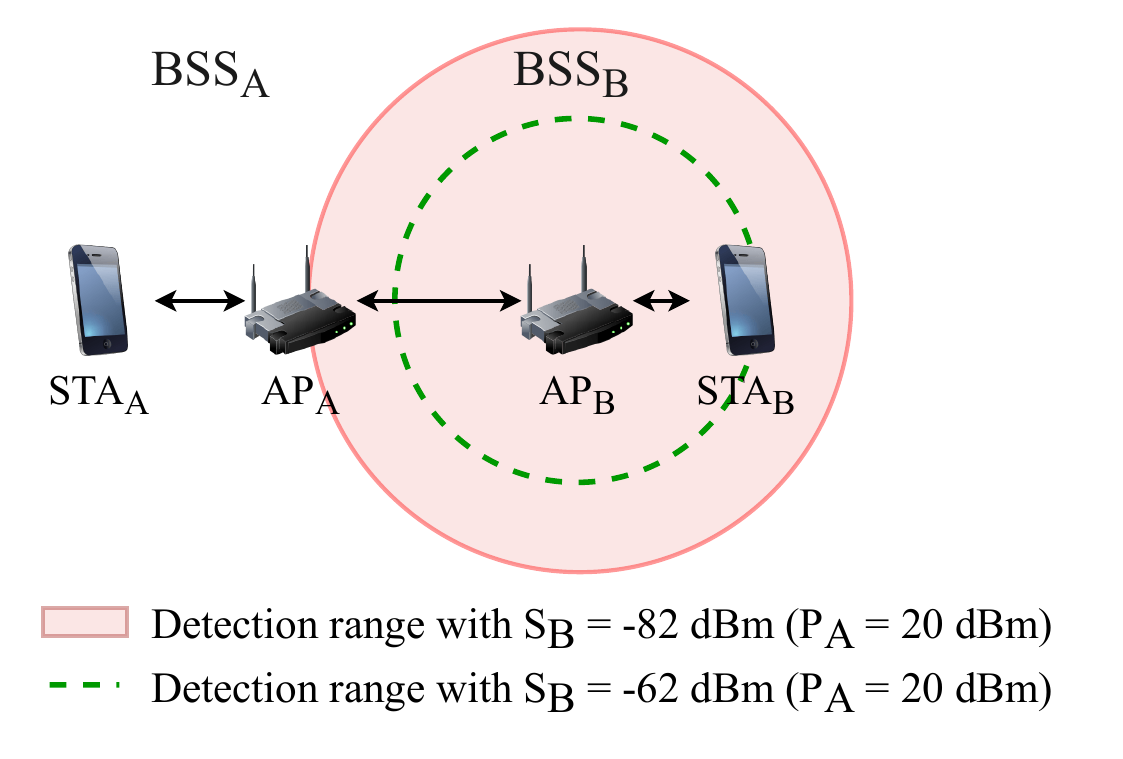}
    \caption{Two \glspl{bss}, each comprising one \gls{ap} and one \gls{sta}, coexist within the same area. The sensitivity range from AP$_B$ indicates whether the transmissions from AP$_A$ are ignored (AP$_A$ is outside the circle) or not (AP$_A$ is inside the circle). P$_X$ and S$_Y$ denote the power and sensitivity used by \gls{ap} $X$ and $Y$, respectively.}
    \label{fig:example_SR}
\end{figure}

In wireless communications, \gls{sr} refers to the ability of wireless devices to perform simultaneous transmissions using the same or overlapping frequency channels. This can be done by properly tuning parameters such as the transmit power $P$ (\textit{determining the interference caused to others}) and \gls{cst} $S$ (\textit{determining the interference tolerance from others})~\cite{wilhelmi2021spatial}. In particular, the transmit power determines the reach of transmissions and the potential quality of the signal at the intended receiver (see Fig.~\ref{fig:example_SR}). A \textit{high power} (e.g. $20$ dBm) would lead to a better \gls{snr}, but would incur higher interference to other devices. A \textit{low power} (e.g., $3$ dBm) would create less interference to other devices, but results in a poorer \gls{snr}. On the other hand, adjusting the listening sensitivity, namely \gls{cst}, in the context of \gls{lbt} (e.g., used for preamble detection) dictates the device's aggressiveness in accessing the medium. Using a low threshold (e.g., $-82$ dBm) would lead to a higher number of transmission deferrals, while a high threshold (e.g., $-62$ dBm) would allow the device to transmit more often (but potentially under higher interference conditions).

\subsection{Spatial Reuse in Wi-Fi}

The implementation of \gls{sr} in Wi-Fi has evolved since the 802.11ax amendment, which first introduced \gls{obsspd}-based \gls{sr}~\cite{wilhelmi2021spatial}. This feature allows devices to ignore inter-\gls{bss} interference by using a \gls{cst} threshold higher than the \gls{cca} for frames that carry a different \gls{bss} Color (i.e., inter-\gls{bss} transmissions). However, to protect any ongoing transmissions, a strict transmit power limitation is enforced during an \gls{sr} \gls{txop} that has been obtained by using a higher \gls{cst}. A few years later, IEEE 802.11be introduced \gls{psr}~\cite{de2020latency}, which facilitates concurrent (uplink) transmissions through \gls{txop} sharing. Similar to \gls{obsspd}, \gls{psr} requires the secondary transmitter to adhere to power limits imposed by the primary \gls{ap} (e.g., based on its tolerated interference). Looking ahead, 802.11bn aims to standardize \gls{cosr}~\cite{wilhelmi2025coordinated}, a natural extension of these mechanisms tailored to the \gls{mapc} framework.

\subsection{Reinforcement Learning Solutions for Spatial Reuse}

To address the \gls{sr} problem, decentralized \gls{rl} has been widely adopted, including works that have applied Q-Learning and \gls{mab} to optimize the transmit power and \gls{cst}~\cite{wilhelmi2017implications, wilhelmi2019collaborative}. In these works, \gls{rl} algorithms are used to maximize individual utilities (e.g., per-\gls{bss} throughput), which proves effective for addressing the inherent complexity of the problem in certain scenarios. However, when applied concurrently by multiple agents, achieving optimal equilibria becomes difficult given the aggressive strategies learned by the agents in such a competitive environment. To address this, other solutions have been proposed, including reward-sharing bandits~\cite{wilhelmi2025coordinated}, cooperative bandits~\cite{iturria2024cooperate}, mechanisms based on interferer identification~\cite{yin2019learning}, and complex wireless state characterization through \gls{drl}~\cite{huang2022deep}.

Non-coordinated strategies, in line with Wi-Fi's decentralized nature, largely rely on rational learning and partial information (e.g., external regret minimization), which trap agents in inefficient equilibria dominated by aggressive strategies (e.g., all the agents use a high transmit power as a protective measure). To overcome the limitations of decentralized learning, Game Theory stands as a promising tool to achieve the high efficiency of coordination without incurring large overheads. In this regard, we find early works proposing cooperative games (e.g., Shapley values, bargaining) in areas like cognitive radio or mesh networking~\cite{song2008joint, bloem2007stackelberg}, focusing primarily on power control and channel allocation. Similarly,~\cite{tan2008game, maghsudi2014channel} provided no-regret solutions for power control in wireless networks. In the context of Wi-Fi, there is a lack of literature applying internal regret minimization strategies, which is precisely the gap this paper aims to fill.

\section{Problem Formulation and Solution Proposal}
\label{sec:formulation}

\subsection{Formulation as an Online Decision-Making Game}

We formulate the optimization of \gls{sr} in Wi-Fi as a \gls{ma} \gls{mab} game, where a set of agents $\mathcal{N} = \{1, \dots, N\}$ operate independently and iteratively select an action from a set $\mathcal{A} = \{1, \dots, K\}$, leading to a joint action profile $\boldsymbol{a} = (a_1, \dots, a_N)$. The game is divided into $T$ slots, where each slot represents a decision-making opportunity for the agents, and the goal of each agent is to optimize its configuration and maximize its own performance (e.g., throughput). In a given iteration $t\in \{1, \dots, T\}$, agents obtain bandit feedback, meaning that they only observe their own reward $r_n$, which depends on the joint strategy profile $r_n^t = f_r(a_n^t, a_{-n}^t)$, where$f_r(\cdot)$ is the reward function and $a_{-n}^t$ denotes the actions selected by all other \glspl{bss}. In the \gls{sr} problem, the action space for each \gls{bss} consists of a discrete pair of transmit power and \gls{cst} values $a = (P, S)$, where $P \in \mathcal{P}$ and $S \in \mathcal{S}$ (with $\mathcal{P}$ and $\mathcal{S}$ being the transmit power and sensitivity action spaces, respectively). The fundamental challenge in this problem is that $r_n$ is non-convex and non-stationary as other \glspl{bss} adapt dynamically. Consequently, standard optimization approaches (e.g., external regret minimization) typically lead to \gls{ne}, which are often inefficient in interference channels~\cite{wilhelmi2019potential}.

\subsection{External vs. Internal Regret}
\label{sec:external_vs_internal}

In Wi-Fi, \gls{sr} interactions are complex due to the consequent contention, collisions, and other adverse effects stemming from decentralization (e.g., unexpected collisions due to hidden nodes, additive interference). Moreover, the bandit (partial) feedback means agents only observe the payoff of the actions they play in each iteration. To define a strategy for individual agents, we can opt for minimizing either the \textit{external} or the \textit{internal} regret~\cite{blum2007external}.

The external regret ($R_n^{\text{ext}}$) looks at the best-performing action in hindsight (i.e., the difference between the received reward and the reward that could have been achieved by playing the single best fixed action over the entire game) to quantify the performance of an action-selection algorithm. In particular, the external regret for \gls{bss} $n$ at time horizon $T$ is
\begin{equation}
    R_n^{\text{ext}}(T) = \max_{k \in \mathcal{A}_n} \sum_{t=1}^T r_n(k, a_{-n}^t) - \sum_{t=1}^T r_n(a_n^t, a_{-n}^t),
\end{equation}
where the first term of the equation refers to the performance of the best fixed strategy, and the second term represents the actual obtained performance. Here, $r_n(k, a_{-n}^t)$ denotes the hypothetical reward the agent would have obtained if it had played action $k$ in iteration $t$, given $a_{-n}^t$.

While external regret minimization works well in static setups, it is often insufficient for multi-agent settings, as it fails to capture the non-stationary dynamics of adaptive opponents (i.e., the external regret looks more at the history of the actions rather than at game dynamics). This typically drives the system to \gls{ne} states characterized by aggressive, selfish behaviors (e.g., employ maximum transmit power) that result in suboptimal network-wide performance. To address this, the internal regret, instead of comparing actions alone, compares the loss of an online algorithm to the loss of a modified online algorithm, which consistently replaces one action with another. This favors the enforcement of \gls{ce}, which, contrary to \gls{ne} (which assumes independent strategies), allows correlating the agents' joint actions, implicitly capturing how the system responds to specific moves. This often leads to higher system-wide efficiency and fairness. In particular, the cumulative internal regret achieved by \gls{bss} $n$ at time $T$ is
\begin{equation}
    R_n^{\text{int}}(T) = \max_{j, k \in \mathcal{A}_n} \sum_{t: a_n^t = j} \left( r_n(k, a_{-n}^t) - r_n(j, a_{-n}^t) \right),
\end{equation}

where the summation is done only for the iterations where action $j$ was played (i.e., $t:a_n^t = j$), and $r_n(k, a_{-n}^t)$ is the hypothetical reward the agent would have obtained if it had played $k$ instead of $j$ in iteration $t$.

\subsection{Solution proposal}

Our proposed solution, detailed in Alg.~\ref{alg:sr_regret}, is based on the independent regret-matching algorithm of~\cite{hart2000simple}. The core idea of regret-matching is that each agent iteratively maintains a cumulative swap-regret matrix $\boldsymbol{Q}$, where each entry $Q_{j \to k}$ includes the estimated cumulative gain of having played action $k$ instead of action $j$. In every iteration $t$, agent $n$ selects the action that maximizes a preference vector $\boldsymbol{\pi}^{t}$, i.e., $a_n^t = \operatorname*{arg\,max}_{k \in \mathcal{A}} , \boldsymbol{\pi}^{t}$ (Alg.~\ref{alg:sr_regret}, line 5). Unlike typical regret-matching, which adopts a mixed strategy based on a probability distribution, here we use a pure strategy to mitigate the negative effects of random exploration and provide better network stability.

At the end of an iteration, each agent receives a reward $r_n^t$ for the played action (Alg.~\ref{alg:sr_regret}, line 6). Crucially, to enable swap-regret minimization, performance estimates (Alg.~\ref{alg:sr_regret}, lines 11-12) are used to update the matrix $\boldsymbol{Q}$ (Alg.~\ref{alg:sr_regret}, line 17), which effectively drives the update of the preference vector $\boldsymbol{\pi}^{t+1}$. Notice that those preferences are updated in proportion to the internal regrets stored in $\boldsymbol{Q}$, allowing for moving away from suboptimal actions and choosing those that offer higher rewards. In addition, a decay factor $\lambda$ is applied to better handle non-stationary and avoid relying on too past information. The normalization of the preferences is done using a parameter $\mu$ (originally used as \emph{inertia} in \cite{hart2000simple}), which ensures that all the values sum to 1 and that no preference value becomes negative (Alg.~\ref{alg:sr_regret}, lines 20-26). For that, $\mu$ is set to $2 (|\mathcal{A}| - 1)$ to ensure that it is strictly greater than the maximum possible sum of positive regrets that an agent could ever accumulate in a single step. 

\begin{algorithm}[t!]
\caption{Regret-matching for decentralized spatial reuse}
\label{alg:sr_regret}
\begin{algorithmic}[1]
\State \textbf{Input:} $\mathcal{A}$ (including power/sensitivity pairs)
\State \textbf{Initialize:} $Q_{j \to k} \leftarrow 0, \forall j, k \in \mathcal{A}$, $\boldsymbol{\pi} \leftarrow \{\frac{1}{|\mathcal{A}|}, ..., \frac{1}{|\mathcal{A}|}\}$, $\mu \leftarrow 2 (|\mathcal{A}| - 1)$ (stickiness factor), $\lambda \leftarrow 0.95$ (decay factor)

\For{$t = 1, \dots, T$}
    \State \textbf{// Action selection \& reward estimation}    
    \State $a^t \leftarrow \operatorname*{arg\,max}_{k \in \mathcal{A}} , \boldsymbol{\pi}^{t}$
    \State $r_{\text{actual}}^t \leftarrow \Gamma^t/\Gamma_{\max}$ (norm. throughput)
    \For{$k \in \mathcal{A}$}
        \If{$k == a^t$}
            \State $\hat{r}_k^t \leftarrow r_{\text{actual}}^t$
        \Else
            \State Estimate airtime $(\hat{\tau}_k)$ \& rate $(\hat{\nu}_k)$
            \State $\hat{r}_k^t \leftarrow \hat{\tau}_k \cdot \hat{\nu}_k$
        \EndIf
    \EndFor   
    \State \textbf{// Update internal regret matrix}
    \For{$k \in \mathcal{A}$}
        \State $Q_{a^t \to k} \leftarrow \max\big(0, \lambda \cdot Q_{a^t \to k} + \left( \hat{r}_k^t - \hat{r}_{a^t}^t \right)\big)$
    \EndFor
    \State \textbf{// Update preference vector}
    
    \For{$k \in \mathcal{A}$}
        \If{$k \neq a^t$}
            \State $\pi_k^{t+1} \leftarrow \frac{1}{\mu} [Q_{a^t \to k}]_+$
        \Else
            \State $\pi_k^{t+1} \leftarrow 1 - \frac{1}{\mu} \sum_{k \neq a^t} [Q_{a^t \to k}]_+$
        \EndIf
    \EndFor
\EndFor
\end{algorithmic}
\end{algorithm}

\textbf{Estimated reward calculation:} 
We define the actual reward of the played action $a$ as the normalized throughput it has experienced, i.e., $r_\text{actual}^t = \Gamma^t/\Gamma_\text{max}$. To estimate the hypothetical reward of the remaining unplayed actions in iteration ($\forall k \neq a^t \in \mathcal{A}$), we propose an estimator that relies on the ``good faith'' of the other agents, thus aiming at achieving \gls{ce}. In particular, we estimate the reward of action $n$ based on its potential performance:
\begin{equation}
    \hat{r}_k^t = \hat{\tau}_k^t \cdot \hat{\nu}_k^t ,
\end{equation}

where $\hat{\tau}_k^t$ estimates the airtime when using sensitivity $S_k$ (\textit{can we ignore the others?}) and power $P_k$ (\textit{are the others ignoring us?}) and $\hat{\nu}(k)$ is the estimated bit rate for action $k$. The latter is extracted from 802.11 tables according to the best \gls{mcs} that supports the estimated \gls{rssi} at the station ($\hat{\text{RSSI}}_{\text{sta}}$).

The estimated airtime captures the non-linear interactions of \gls{csmaca}, specifically addressing contention, starvation (fairness), and the capture effect. It is computed as
\begin{equation}\hat{\tau}_k = \frac{\eta}{\psi_\text{cont} \cdot \psi_{\text{fair}}} ,
\end{equation}

where $\psi_\text{cont} = 1 + \sum_{m \in \mathcal{N}} \mathbb{I}(\hat{\text{RSSI}}_m \geq S_k)$ is a contention term that represents the number of devices sharing the medium (which depends on the estimated \gls{rssi} from node $m\neq n$). $\psi_\text{fair}$ is a fairness penalty used to prevent hidden nodes. It is set to $\omega$ if starvation is detected due to asymmetric interactions, i.e., when agent $n$ cannot detect neighbor $m$ ($\hat{\text{RSSI}}_m < S_k$) but in turn transmits with enough power to silence it ($\hat{\text{RSSI}}_n \geq {\text{CCA}}$), the agent is causing starvation. Otherwise, it is set to 1. Finally, $\eta = \mathbb{I}(\hat{\gamma_k} > \text{CE})$ ensures that the estimated \gls{sinr} based on the power used by the others does not fall below the capture effect threshold (which would potentially lead to packet losses). 

The behavior of the considered algorithm is tightly coupled with the accuracy of the reward estimator, provided that the cumulative regret matrix $\boldsymbol{Q}$ is based on the difference between the real revealed reward $r_{\text{actual}}$ and the hypothetical estimates $\hat{r}_k$. Specifically, underestimating the potential of unplayed actions may still trap the system in a suboptimal equilibrium. Conversely, overestimating rewards encourages exploration, potentially inducing instability and unfairness. Nevertheless, the estimator does not need to be perfect, as its primary goal is to serve as a heuristic for agents, filling the gap between complete decentralization and centralization.

\section{Performance Evaluation}
\label{sec:performance_evaluation}

\begin{figure}[t!]
    \centering
    \includegraphics[width=.75\columnwidth]{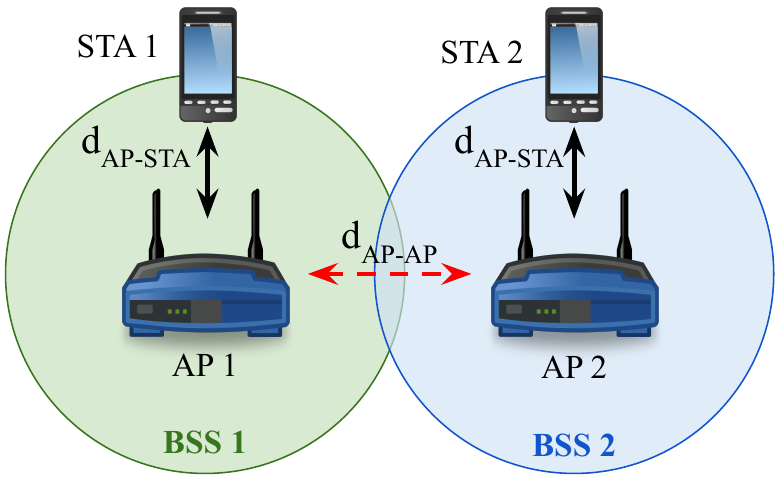}
    \caption{Considered 2-BSS scenario.}
    \label{fig:scenario}
\end{figure}

The evaluation of the proposed algorithm is performed using Komondor~\cite{barrachina2019komondor},\footnote{For the sake of openness and reproducibility, all the source code used in this paper is open and can be accessed at \url{https://github.com/wn-upf/Komondor/tree/internal_regret_minimization} (commit: 0fb1be3).} a Wi-Fi simulator that includes \gls{ml} agents for driving the optimization of various features, including \gls{sr}. We consider a scenario comprising two \glspl{bss}, each including an \gls{ap} and a \gls{sta} (see Fig.~\ref{fig:scenario}). This scenario allows for a more precise study of game-theoretic phenomena, given that the interactions between the two players can be easily analyzed. Furthermore, this fits 11bn's \gls{mapc}, which considers coordinated transmissions from two \glspl{ap} only. More \glspl{bss} will be considered in our future work.

Specifically, we compare three different approaches that both agents can take:
\begin{itemize}
    \item Baseline (\texttt{static}): Standard 802.11 operation with fixed parameters ($P=20$ dBm, $S=-82$ dBm).
    \item External Regret (\texttt{$\varepsilon$-greedy}): A standard \gls{mab} approach ($\varepsilon$-greedy) that seeks to maximize individual reward, converging to \gls{ne}. In $\varepsilon$-greedy, each agent selects a random action with probability $\varepsilon$, while the rest of the times ($1-\varepsilon$) exploits the action that has shown the highest payoff so far.
    \item Internal Regret (\texttt{regret-matching}): The proposed regret-matching algorithm described in Section~\ref{sec:formulation}, converging to \gls{ce}.
\end{itemize}

The main simulation parameters are included in Table~\ref{tab:simulation_parameters}.

\begin{table}[t!]
\centering
\caption{Simulation parameters.}
\label{tab:simulation_parameters}
\resizebox{\columnwidth}{!}{%
\begin{tabular}{@{}clc@{}}
\toprule
Parameter & \multicolumn{1}{c}{\textbf{Description}} & \textbf{Value} \\ \midrule
$T_\text{sim}$ & Simulation time & $100$ s \\
$S$ & Number of simulated random deployments & 100 \\
$F_c$ & Carrier frequency & $5$ GHz \\
$\text{GI}$ & Guard Interval & $3.2$ $\mu$s\\ 
$B$ & Transmission bandwidth & $20$ MHz \\
$\text{MCS}$ & MCS indices & 0-11\\ 
$\mathcal{P}^\text{Noise}$ & Noise power & $-95$ dBm \\
$\mathcal{P}_{tx,\max}$ & Default transmit power & $20$ dBm \\
CCA & Default CCA threshold & $-82$ dBm \\
$N_\text{ss}$ & Single-user spatial streams & $1$ \\
$G^\text{TX/RX}$ & Transmitter/receiver antenna gain & $0/0$ dBi \\
$\textrm{CE}$ & Capture effect threshold & $10$ dB \\
PL & Path loss model and parameters & See \cite{wilhelmi2019potential} \\
$\text{TXOP}_{\max}$ & TXOP duration limit & $5.484$ ms\\ 
$\text{A-MPDU}_{\max}$ & A-MPDU size & $64$\\ 
$L_{D}$ & Length of data packets & $1500$ bytes \\ 
$\mathcal{T}$ & Traffic model & Full-buffer\\ 
$\mathbb{T}$ & Traffic type & Downlink (DL) \\ 
$\text{CW}_0$ & Initial Contention Window (CW) & $16$ \\
$\text{CWE}_{\min/\max}$ & Min./Max. CW exponent & $1/5$ \\
\midrule
$\mathcal{S}$ & CST values & $\{-62, -72, -82\}$~dBm\\
$\zeta$ & Transmit power values & $\{5, 10, 15, 20\}$~dBm\\
$\Delta$ & Iteration duration & $0.5$~s \\
$\varepsilon_0$ & Initial exploration coefficient & $0.1$ \\
$\varepsilon(t)$ & Exploration coefficient & $\varepsilon_0/\sqrt{t}$ \\
$\omega$ & Fairness penalty & $4 (2 \cdot N)$ \\
$\lambda$ & Discount factor & $0.95$ \\
\bottomrule
\end{tabular}%
}
\end{table}

\subsection{Toy Scenarios}

To understand the behavior of agents using external or internal regret minimization strategies, we start with concrete configurations of the deployment depicted in Fig.~\ref{fig:scenario}, which allows the definition of two different learning problems:\footnote{For the sake of analysis, in this setup we limit the actions to two different values of sensitivity (-72 dBm, -82 dBm) and power (10 dBm, 20 dBm).} $i)$ \textit{strong equilibrium} (for $d_\text{AP-AP} = 5$~m and $d_\text{AP-STA} = 2$~m) and $ii)$ \textit{weak equilibrium} (for $d_\text{AP-AP} = 4$~m and $d_\text{AP-STA} = 2$~m). Varying the positions of the devices leads to different spatial interactions among \glspl{bss}. In the first case (strong equilibrium), both \glspl{ap} can successfully transmit in parallel if they increase their sensitivity (e.g., using $S=-72$~dBm and $P=20$~dBm). Therefore, the strategy maximizing the overall performance is the same as for maximizing the individual performance. As a result, finding the optimal solution is straightforward for greedy agents. This does not occur in the weak-equilibrium case, where simultaneous transmissions at the maximum transmit power would lead to collisions at both \glspl{sta}. In that case, the optimal action consists of lowering the transmit power to minimize the interference generated towards the other \gls{bss} ($P=10$~dBm) and increasing the \gls{cst} to avoid contention ($S=-72$~dBm).

\begin{figure}[t!]
    \centering
    \includegraphics[width=\columnwidth]{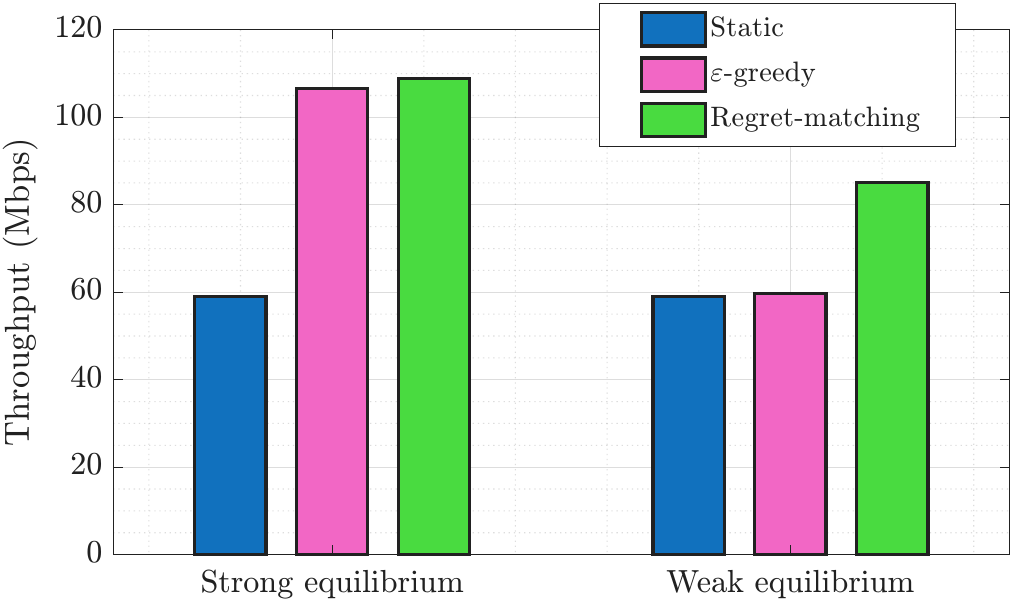}
    \caption{Mean average throughput (in Mbps) obtained by the two BSSs in the scenario encompassing strong and weak equilibrium situations.}
    \label{fig:throughput_toy_scenario}
\end{figure}

The mean average throughput achieved across the two \glspl{bss} during the simulations of the two cases (strong vs. weak equilibrium) is shown in Fig.~\ref{fig:throughput_toy_scenario} for each considered approach. As shown, in the first case (strong equilibrium), the static (default) configuration leads to an average throughput of $\sim$60~Mbps because the two \glspl{bss} alternate the access to the medium. In that case, both the external ($\varepsilon$-greedy) and internal (regret-matching) regret minimization approaches allow reusing the space and performing simultaneous transmissions. It is worth noting that, because it sticks to the best action and performs to further exploration, regret-matching leads to slightly higher performance than $\varepsilon$-greedy, which in turn continues to explore random actions even after finding the best action set. When it comes to the weak equilibrium case, we find that the $\varepsilon$-greedy approach gets stuck in the default performance, provided that the agents play aggressively and are not capable of finding the best configuration. In turn, when using regret-matching, the agents are able to discover the best-performing action, thus leading to a mean average throughput above $80$~Mbps.

\begin{figure}[t!]
\centering
   \begin{subfigure}{0.23\textwidth}
        \centering
        \includegraphics[width=\linewidth]{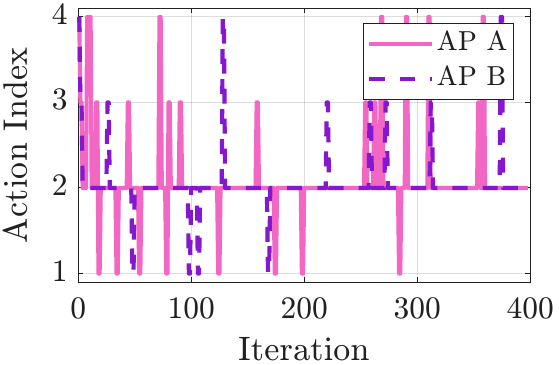}
        \caption{Strong eq. ($\varepsilon$-greedy)}
        \label{fig:2a}
   \end{subfigure}
   \hfill 
   \begin{subfigure}{0.23\textwidth}
        \centering
        \includegraphics[width=\linewidth]{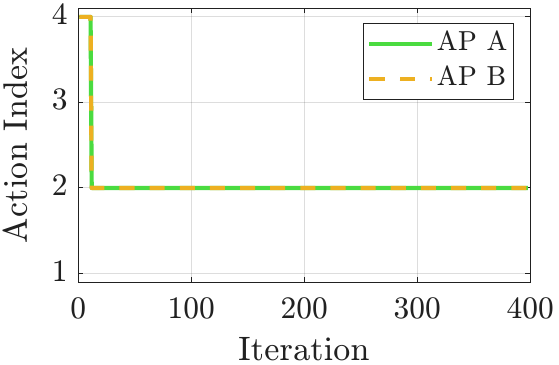}
        \caption{Strong eq. (regret-matching)}
        \label{fig:2b}
   \end{subfigure}
   \hfill
   \begin{subfigure}{0.23\textwidth}
        \centering
        \includegraphics[width=\linewidth]{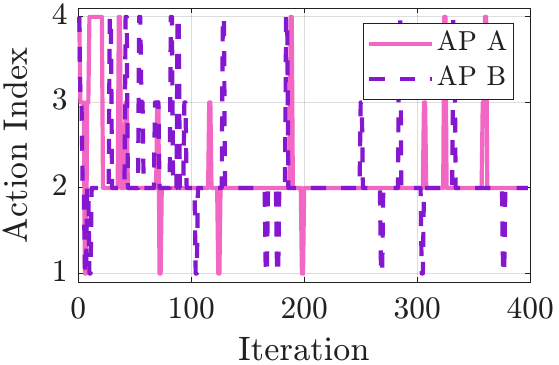}
        \caption{Weak eq. ($\varepsilon$-greedy)}
        \label{fig:2c}
   \end{subfigure}
   \hfill
   \begin{subfigure}{0.23\textwidth}
        \centering
        \includegraphics[width=\linewidth]{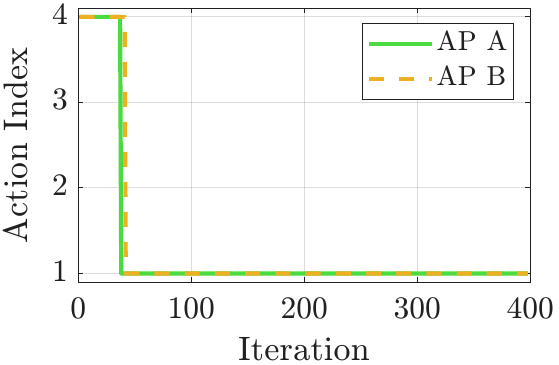}
        \caption{Weak eq. (regret-matching)}
        \label{fig:2d}
   \end{subfigure}
   
   \caption{Actions selected by the two BSSs per iteration in each scenario. Actions available (CST, transmit power): $A_1$ = \{-72, 10\} dBm, $A_2$ = \{-72, 20\} dBm, $A_3$ = \{-82, 10\} dBm, $A_4$ = \{-82, 20\} dBm.}
   \label{fig:actions_selected}
\end{figure}

To better understand the behavior of each learning algorithm, Fig.~\ref{fig:actions_selected} shows the actions selected by each agent in every iteration. Starting with the strong equilibrium scenario, $\varepsilon$-greedy (Fig.~\ref{fig:2a}) discovers the best action ($A_2$), but continues exploring suboptimal actions throughout the simulation. Regret-matching (Fig.~\ref{fig:2b}), in contrast, quickly converges to the best action and does not perform any other exploration. In this case, regret-matching spends several iterations stuck in a suboptimal arm ($A_4$), but soon switches to the true optimal arm as a result of having accumulated positive regret for such an action during several consecutive iterations. When it comes to the weak equilibrium case, $\varepsilon$-greedy gets stuck in a suboptimal arm (see Fig.~\ref{fig:2c}), whereas regret-matching quickly converges to the optimal one (see Fig.~\ref{fig:2d}).

\subsection{Random Deployments}

\begin{figure}[t!]
    \centering
    \includegraphics[width=\columnwidth]{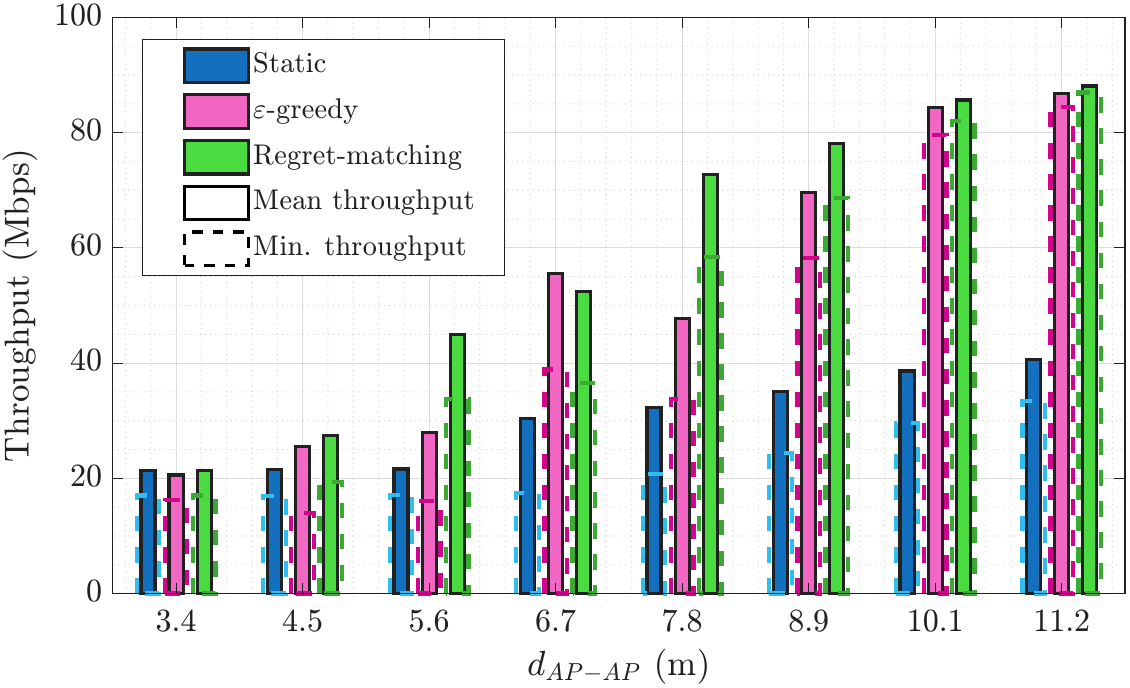}
    \caption{Mean (solid bars) and minimum (dashed bars) average throughput achieved across random deployments.}
    \label{fig:fig2_random_scenarios}
\end{figure}

To showcase the potential of the regret-matching solution, we now consider random deployments comprising the two studied \glspl{bss} from Fig.~\ref{fig:scenario}. In particular, we study the effect of inter-\gls{ap} interference, accounting for different distances between \glspl{bss} based ($d_\text{AP-AP}$). The position of the \glspl{sta} is also randomly selected around their \glspl{ap}, with $d_\text{AP-STA} = [3-5]$~m. The throughput results (mean and minimum across \glspl{bss}) are depicted in Fig.~\ref{fig:fig2_random_scenarios}.

As shown, regret-matching consistently improves the performance of the two \glspl{bss} across different deployment setups. For shorter distances, increasing \gls{sr} is challenging, hence the gains are limited. However, as the distance between \glspl{bss} increases, hence the inter-\gls{ap} interference decreases, the \gls{sr} performance gains are more significant. When it comes to the minimum throughput, which embodies fairness in all \glspl{bss}, regret-matching offers better performance in most cases, thus underscoring its game-theoretic design.

\section{Conclusions}
\label{sec:conclusions}

As Wi-Fi networks evolve towards the upcoming IEEE 802.11bn (Wi-Fi 8), the management of inter-\gls{bss} interference has become an important challenge for reliability. While the standard is shifting toward centralization with \gls{mapc} (i.e., \gls{cosr}), such architectures impose significant signaling overheads that increase complexity and limit scalability. In this paper, we proposed a learning-based method that leverages Game Theory to bridge the gap between decentralized decision-making and global optimality. In particular, our method adopts an internal regret minimization approach (based on regret-matching) to achieve \gls{ce}. With this, we aimed to achieve implicit coordination among \glspl{bss}, but without having to exchange a single bit for coordination purposes. Our simulation results confirm that this approach has practical potential, as it has been shown to overcome the limitations of state-of-the-art external regret minimization methods. As the IEEE 802.11 groups define the specifications for next-generation Wi-Fi networks, algorithms grounded in internal regret minimization offer a compelling alternative to scalable, high-efficiency, and flexible optimization solutions, unlocking the full potential of the unlicensed spectrum.

\section*{Acknowledgments}
This paper is supported by the CHIST-ERA Wireless AI 2022 call MLDR project (ANR-23-CHR4-0005), partially funded by AEI under project PCI2023-145958-2, by TRUE-Wi-Fi PID2024-155470NB-I00 and Wi-XR PID2021-123995NB-I00 (MCIU/AEI/FEDER,UE), by MCIN/AEI under the Maria de Maeztu Units of Excellence Programme (CEX2021-001195-M), and AGAUR ICREA Academia 00077.
    
\bibliographystyle{IEEEtran}
\bibliography{references}

\end{document}

%% file: acronyms.tex
\newacronym{ieee}{IEEE}{Institute of Electrical and Electronics Engineers}
\newacronym{vr}{VR}{Virtual Reality}
\newacronym{ar}{AR}{Augmented Reality}
\newacronym{obss}{OBSS}{Overlapping Basic Service Set}
\newacronym{iot}{IoT}{Internet of Things}
\newacronym[plural=APs]{ap}{AP}{Access Point}
\newacronym{csmaca}{CSMA/CA}{Carrier Sense Multiple Access with Collision Avoidance}
\newacronym{lbt}{LBT}{Listen-Before-Talk}
\newacronym[plural=BSSs]{bss}{BSS}{Basic Service Set}
\newacronym{mlo}{MLO}{Multi-Link Operation}
\newacronym{sr}{SR}{Spatial Reuse}
\newacronym{obsspd}{OBSS/PD}{Overlapping Basic Service Set/Packet Detect}
\newacronym[plural=STAs]{sta}{STA}{Station}
\newacronym{snr}{SNR}{Signal-to-Noise-Ratio}
\newacronym{tgbn}{TGbn}{Task Group bn}
\newacronym{cosr}{Co-SR}{Coordinated Spatial Reuse}
\newacronym{cobf}{Co-BF}{Coordinated Beamforming}
\newacronym{cortwt}{Co-RTWT}{Coordinated R-TWT}
\newacronym{cotdma}{Co-TDMA}{Coordinated Time Division Multiple Access}
\newacronym{jt}{JT}{Joint Transmissions}
\newacronym{coofdma}{Co-OFDMA}{Coordinated OFDMA}
\newacronym{mapc}{MAPC}{Multi-Access Point Coordination}
\newacronym{txop}{TXOP}{Transmission Opportunity}
\newacronym{ecsr}{ECSR}{Enhanced Coordinated Spatial Reuse}
\newacronym{rssi}{RSSI}{Received Signal Strength Indicator}
\newacronym{mcs}{MCS}{Modulation and Coding Scheme}
\newacronym{dcf}{DCF}{Distributed Coordination Function}
\newacronym{sinr}{SINR}{Signal-to-Interference-plus-Noise Ratio}
\newacronym{rrm}{RRM}{Radio Resource Management}
\newacronym{rts}{RTS}{Ready-to-Send}
\newacronym{cts}{CTS}{Clear-to-Send}
\newacronym{arq}{ARQ}{Automatic Repeat Request}
\newacronym{ampdu}{A-MPDU}{Aggregate MAC Protocol Data Unit}
\newacronym{ai}{AI}{Artificial Intelligence}
\newacronym{ml}{ML}{Machine Learning}
\newacronym{phy}{PHY}{Physical}
\newacronym{mac}{MAC}{Medium Access Control}
\newacronym{uhr}{UHR}{Ultra High Reliability}
\newacronym{sifs}{SIFS}{Short Interframe Space}
\newacronym{difs}{DIFS}{DCF Interframe Space}
\newacronym{aifs}{AIFS}{Arbitration Interframe Space}
\newacronym{wds}{WDS}{Wireless Distribution System}
\newacronym{ssid}{SSID}{Service Set Identifier}
\newacronym{ess}{ESS}{Extended Service Set}
\newacronym{mbss}{MBSS}{Mesh Basic Service Set}
\newacronym{wlc}{WLC}{Wireless LAN Controller}
\newacronym{wlan}{WLAN}{Wireless Local Area Network}
\newacronym{capwap}{CAPWAP}{Control And Provisioning of Wireless Access Points}
\newacronym{ietf}{IETF}{Internet Engineering Task Force}
\newacronym{rfc}{RFC}{Request for Comments}
\newacronym{isp}{ISP}{Internet Service Provider}
\newacronym{cpe}{CPE}{Customer-Premises Equipment}
\newacronym{usp}{USP}{User Services Platform}
\newacronym{mqtt}{MQTT}{Message Queuing Telemetry Transport}
\newacronym{bssid}{BSSID}{Basic Service Set Identifier}
\newacronym{tpc}{TPC}{Transmit Power Control}
\newacronym{cocr}{Co-CR}{Coordinated Channel Recommendation}
\newacronym{pasn}{PASN}{Pre-Association Security Negotiation}
\newacronym{ltf}{LTF}{Long Training Field}
\newacronym{gi}{GI}{Guard Interval}
\newacronym{twt}{TWT}{Target Wake Time}
\newacronym{sp}{SP}{Service Period}
\newacronym{rtwt}{R-TWT}{Restricted Target Wake Time}
\newacronym{aid}{AID}{Association ID}
\newacronym{pmksa}{PMKSA}{Pairwise Master Key Security Association}
\newacronym{ptksa}{PTKSA}{Pairwise Transient Key Security Association}
\newacronym{rsne}{RSNE}{Robust Security Network Element}
\newacronym{rsnxe}{RSNXE}{RSN eXtension Element}
\newacronym{mic}{MIC}{Message Integrity Check}
\newacronym{back}{BACK}{Block Acknowledgment}
\newacronym{ppdu}{PPDU}{Physical Layer Protocol Data Unit}
\newacronym{csi}{CSI}{Channel State Information}
\newacronym{ndp}{NDP}{Null Data Packet}
\newacronym{bsrp}{BSRP}{Buffer Status Report Poll}
\newacronym{ntb}{NTB}{Non-Trigger Based}
\newacronym{icf}{ICF}{Initial Control Frame}
\newacronym{icr}{ICR}{Initial Control Response}
\newacronym{dps}{DPS}{Dynamic Power Save}
\newacronym{emlsr}{EMLSR}{Enhanced Multi-Link Single Radio}
\newacronym{ofdm}{OFDM}{Orthogonal Frequency-Division Multiplexing}
\newacronym{eht}{EHT}{Extremely High Throughput}
\newacronym{ndpa}{NDPA}{NDP Announcement}
\newacronym{tb}{TB}{Trigger-Based}
\newacronym{bfrp}{BFRP}{Beamforming Report Poll}
\newacronym{ofdma}{OFDMA}{Orthogonal Frequency-Division Multiple Access}
\newacronym{mu}{MU}{Multi-User}
\newacronym{txs}{TXS}{Triggered TXOP Sharing}
\newacronym{mpdu}{MPDU}{MAC Protocol Data Unit}
\newacronym{tsf}{TSF}{Time Synchronization Function}
\newacronym{p2p}{P2P}{Peer-to-Peer}
\newacronym{txspg}{TXSPG}{TXOP Sharing for P2P Groups}
\newacronym{scs}{SCS}{Stream Classification Service}
\newacronym{qos}{QoS}{Quality of Service}
\newacronym{ru}{RU}{Resource Unit}
\newacronym{mimo}{MIMO}{Multiple Input Multiple Output}
\newacronym{cst}{CST}{Carrier Sense Threshold}
\newacronym{irm}{IRM}{Internal Regret Minimization}
\newacronym{ce}{CE}{Correlated Equilibrium}
\newacronym{ne}{NE}{Nash Equilibrium}
\newacronym{cca}{CCA}{Clear Channel Assessment}
\newacronym{psr}{PSR}{Parametrized Spatial Reuse}
\newacronym{mab}{MAB}{Multi-Armed Bandit}
\newacronym{ma}{MA}{Multi-Agent}
\newacronym{drl}{DRL}{Deep Reinforcement Learning}
\newacronym{rl}{RL}{Reinforcement Learning}